\documentclass[preprint,showpacs,preprintnumbers,amsmath,amssymb,nofootinbib]{revtex4}

\usepackage{amsthm}

{Corollary}[section]
{Proposition}[section]

\usepackage{mathrsfs}
\usepackage{etex}
\usepackage{amssymb,amsthm,amscd,amsbsy,array}
\usepackage{bm}
\usepackage{amsmath,dsfont}
\usepackage{soul} 
\usepackage{graphics,graphicx,xcolor}

\usepackage[colorlinks=true, pdfstartview=FitV, linkcolor=blue, citecolor=blue, urlcolor=blue]{hyperref} 

\newcommand{\red}{\textcolor{red}}

\newcommand{\blue}{\textcolor{blue}}

\newcommand{\gb}{\quad\colorbox{green}}

\newcommand{\dgreen}{\textcolor[rgb]{0,0.5,0}}

\newenvironment{redtext}{\color{red}}
{\ignorespacesafterend}
\newenvironment{bluetext}{\color{blue}}{\ignorespacesafterend}
\newenvironment{greentext}{\color{green}}{\ignorespacesafterend}
\newenvironment{magentatext}{\color{magenta}}{\ignorespacesafterend}
\newenvironment{cyantext}{\color{cyan}}{\ignorespacesafterend}
\newenvironment{orangetext}{\color{orange}}
{\ignorespacesafterend}

\newcommand{\bmagenta}{\begin{magentatext}}
\newcommand{\emagenta}{\end{magentatext}}
\newcommand{\bcyan}{\begin{cyantext}}
\newcommand{\ecyan}{\end{cyantext}}
\newcommand{\bblue}{\begin{bluetext}}
\newcommand{\eblue}{\end{bluetext}}
\newcommand{\bred}{\begin{redtext}}
\newcommand{\ered}{\end{redtext}}

\newcommand{\bgreen}{\begin{greentext}}
\newcommand{\egreen}{\end{greentext}}
\newcommand{\borange}{\begin{orangetext}}
\newcommand{\eorange}{\end{orangetext}}

\newcommand{\beq}{\begin{equation}}
\newcommand{\eeq}{\end{equation}}
\newcommand{\bec}{\begin{center}}
\newcommand{\ec}{\end{center}}

\DeclareMathOperator{\cn}{cn}
 
\newcommand{\PT}{{P\"oschl{\strut}-Teller\;}}

\newcommand{\GW}{{gravitational wave\,}}

\newcommand{\cA}{{\mathcal{A}}}

\newcommand{\bX}{{\bm{X}}}

\def\aand{{\quad\text{\small and}\quad}}
\def\where{{\quad\text{\small where}\quad}}

\newcommand{\sign}{\mathrm{sign}}

\def\smallover\#1/\#2{\hbox{$\textstyle\frac{\#1}{\#2}$}} %

\def\bequ{\begin{enumerate}}
\def\eenu{\end{enumerate}}
\def\bitem{\begin{itemize}}
\def\eitem{\end{itemize}}

\def\beq{\begin{equation}}
\def\eeq{\end{equation}}
\def\beqa{\begin{eqnarray}}
\def\eeqa{\end{eqnarray}}

\def\barray{\left(\begin{array}}
\def\earray{\end{array}\right)}
\def\barraynb{\begin{array}}
\def\earraynb{\end{array}}

\def\GW{{gravitational wave\;}}
\def\GWs{{gravitational waves\;}}

\def\?{{\,\gb{\fbox{\texttt{??}}\;}}\,}

\def\benu{\begin{enumerate}}
\def\eenu{\end{enumerate}}
\def\bitem{\begin{itemize}}
\def\eitem{\end{itemize}}

\def\smallover#1/#2{\hbox{$\textstyle\frac{#1}{#2}$}} %
\def\smallcirc{{\raise 0.5pt \hbox{$\scriptstyle\circ$}}}
\def\cabove(#1){\stackrel{\smallcirc}{#1}}
\def\ccabove(#1){\,\stackrel{\smallcirc\smallcirc}{#1}\,}
\def\cccabove(#1){\stackrel{\,\smallcirc\smallcirc\smallcirc}{#1}\,}
\def\2{{\smallover1/2}}

\def\cA{{\cal A}}
\def\boxit#1{
\vbox{\hrule\hbox{\vrule\kern4pt
\vbox{\kern5pt#1\kern5pt}\kern4pt\vrule}\hrule}
} 

\def\besub{\begin{subequations}}
\def\esub{\end{subequations}}

\begin{document}

\preprint{\texttt{arXiv: 2502.01326v5 [gr-qc]}}

\title{Flyby-induced displacement effect: an analytic solution}

\author{
P.-M. Zhang$^{1}$\footnote{corresponding author.   zhangpm5@mail.sysu.edu.cn. 
 },
Z.~K.~Silagadze$^{2}$\footnote{silagadze@inp.nsk.su.  
},
and P.~A. Horvathy$^{3}$\footnote{horvathy@univ-tours.fr.}
}

\affiliation{
${}^1$School of Physics and Astronomy, Sun Yat-sen University, Zhuhai 519082, (China)
\\
${}^2$ Budker Institute of Nuclear Physics and Novosibirsk State University, 630 090, Novosibirsk, (Russia).   
\\
${}^{3}$ Institut Denis-Poisson CNRS/UMR 7013 - Universit\'e de Tours - Universit\'e d'Orl\'eans Parc de Grammont, 37200; Tours, (France).
\\
}
\date{\today}


\begin{abstract}
The motion of particles hit by a burst of gravitational waves generated by flyby admits, for the derivative-of-the-Gaussian profile, only a numerical description. The profile can however be approximated by the hyperbolic Scarf potential which admits an exact analytic solution via the Nikiforov-Uvarov method. 
Our toy model is consistent with the prediction of Zel'dovich and Polnarev provided the wave amplitude takes certain ``magical'' values.

\bigskip
Phys. Lett. \textbf{B} 868 (2025) 139687.\,
https://doi.org/10.1016/j.physletb.2025.139687

\bigskip

\end{abstract}
\maketitle

\tableofcontents
\goodbreak

\section{Introduction}\label{Intro}

One of the early proposals to observe \GWs\!,  put forward by Braginsky and Thorne \cite{BraTho}, was to study the motion of particles initially in rest after the passing of a burst of \GWs\!. They argued indeed  that {\sl Burst With Memory could be among the earliest kinds of gravitational waves detected} \cite{BraTho}. Since then, and with the discovery that gravitational memory effects are closely related to soft graviton theorems and Bondi-Metzner-Sachs symmetries at future null infinity, the field has grown considerably \cite{BMS1,BMS2,BMS3,BMS4}.

Initial studies \cite{Ehlers,GWGNature} considered the \emph{Velocity Memory} (VM) effect~: after the wave has passed, the particles would fly apart with constant velocity along straight trajectories.
 Zel'dovich and Polnarev \cite{ZelPol} advocated instead in favor of \emph{flyby}: gravitational waves generated by a cluster of superdense stars (neutron stars or black holes)  would yield the \emph{Displacement  Memory} (DM)  effect~: 
\textit{\narrower  ``\dots
 although the distance between a pair of  
  two noninteracting bodies (such as satellites)
  will change, their relative velocity will become vanishingly small as the flyby concludes."
 }

The GW memory effect offers a way to test general relativity including the nonlinear nature of gravity, and provides additional information about the GW source. Preliminary studies have suggested that the detection of GW memory from individual mergers of massive binary black holes is expected with LISA \cite{Favata}. See \cite{Blanchet} for a state-of-the-art review. However, more realistic estimates show that observing the GW memory effect at detectors such as LISA, although feasible, is a challenging task \cite{fly-1,fly-2}. Recent attempts to detect the effect in LIGO/Virgo's first transient gravitational-wave catalog, as well as in a pulsar timing array experiment NANOGrav, have, predictably, yielded negative results \cite{fly-3,fly-4}. However the prospects for the next generation of gravitational wave detectors are brighter even for future ground-based detectors such as Cosmic Explorer and the Einstein Telescope (see \cite{fly-2,fly-4} and references therein). 

 
Although the displacement effect could  be confirmed  \cite{Christo,BMS1,BMS2,BMS3,BMS4}, there remained some doubts due to the apparent contradiction with the Bondi-Pirani theorem on the caustic property of plane gravitational waves \cite{BoPi89,ShortMemory}. 

More recently, the DM effect could be verified  numerically for plane waves   using the approach proposed in \cite{LongMemory},   provided  the amplitude takes some ``magical" values \cite{DM-1,Jibril,DM-2}.
In this note we study the Zel'dovich - Polnarev  statement by using the simplified framework in ref \cite{LongMemory}. Although they should be viewed only as toy models, they do provide some insight into the memory effect.  

Based Gibbons and Hawking's earlier work \cite{Gibbons_1976}, the  model in \cite{LongMemory} proposes that the Brinkmann profile of the \GW generated by flyby  be approximated by the \emph{first derivative of the Gaussian} (dG), eqn. \eqref{dGprof} below. This yields however Sturm-Liouville-type geodesic equations with no analytic solution. In this paper we propose a similar but slightly different  approximation, namely by the Scarf potential  \cite{Scarf,Scarfpot} which does provide us with analytic solutions. 
The results are in full agreement with \cite{ZelPol}. 

In Brinkmann coordinates, the metric is written as \cite{Brinkmann},
\beq
ds^2=d\bX^2+2dUdV+\cA(U)\big(X_{+}^2-X_{-}^2\big)dU^2\,,
\label{Bmetric}
\eeq
where $\bX$ is referred to as transverse coordinate; $U$ and $V$ are lightlike. In the  non-relativistic Kaluza-Klein-type framework of ref. \cite{DBKP,DGH91} $U$ plays the role of  non-relativistic time.
Ref. \cite{LongMemory}  argues that in linear theory the Riemann tensor of a gravitational wave is proportional to the fourth time derivative of the quadrupole moment of the source \cite{Gibbons_1976,GibbonsP}. This lead us to propose the derivative of the Gaussian,
\beq
\cA^{dG} \propto \frac{\;d}{dU}\Big(\exp[-U^2]\Big)\,
\label{dGprof}
\eeq 
as profile. 
The geodesic equations for \eqref{dGprof} have no analytic solution, though, the profile can however be  approximated by the \emph{derived P\"oschl-Teller} (dPT) potential \cite{PTeller,Chakra,DM-1},
\begin{equation}
    \cA^{dPT}(U)= \frac{\; d}{dU}\left(\frac{g}{2\cosh^{2}U}\right)=-g\frac{\sinh{U}}{\cosh^3{U}}\,.
    \label{dPTprof}
\end{equation} 
Then the geodesics can be studied semi-analytically
\cite{DM-1,DM-2}. 
Our investigations led us to conjecture  
that \emph{DM arises when a quantization} holds, namely when \emph{the wave zone contains an integer number}  of what we called \emph{half-waves}. 
 
\begin{figure}[h]
\includegraphics[scale=.35]{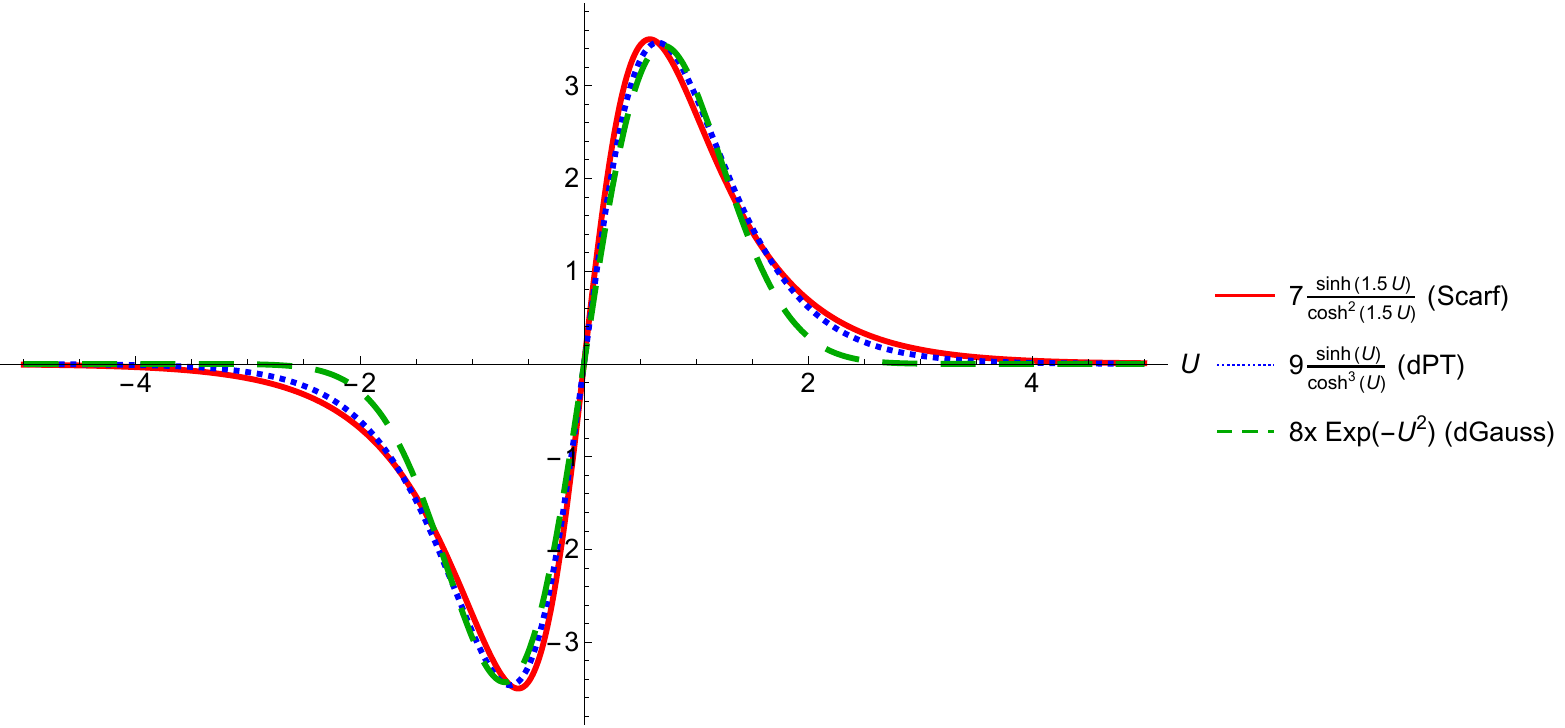}
\vskip-4mm
\caption{\textit{\small The  derived \blue{\PT} profile \eqref{dPTprof} and its \red{Scarf potential}  alter-ego 
\eqref{Scarfpot} are both approximations of the \dgreen{derived-Gaussian} one proposed in  \cite{LongMemory} to describe flyby.} 
\label{ScarfProfFig} 
}
\end{figure}

In this note we  approximate the flyby profile \eqref{dGprof} by the hyperbolic Scarf potential \cite{Scarfpot} \footnote{The trigonometric Scarf potential was initially introduced in \cite{Scarf} to mimic the potential of a real crystal. It turned out to be the tip of the iceberg of a whole class of potentials with interesting properties, ranging from classical and quantum integrability \cite{Perelomov,Perelomov1} to supersymmetric quantum mechanics \cite{Gendenshtein, Natanson,Dutt86,Levai,Scrf1,MSUSY}. },
\begin{equation}
    \cA^{Scarf}(U)=2g\frac{\;\;d}{dU} \left(\frac{1}{\cosh U}\right) =-2g\frac{\sinh{U}}{\cosh^2{U}}\,,   
    \label{Scarfpot}
\end{equation}
to be compared with \eqref{dPTprof}. The difference is merely the exponent in the denominator.
For small values of $U$ the Scarf profile \eqref{Scarfpot} behaves as the derivative of the Gaussian or of P\"oschl-Teller, \eqref{dGprof} or \eqref{dPTprof}, but for large $U$ it falls off more slowly. However dilating $U$ appropriately almost perfect overlapping is obtained,
 as depicted in FIG.\ref{ScarfProfFig}. Their similarity follows from that all three of them are derivatives of similar bell-shaped functions.
 We note also all three profiles  in
\eqref{dGprof}-\eqref{dPTprof}-\eqref{Scarfpot}
 are derivatives of an even function and are therefore odd.

The geodesics are, in Brinkmann coordinates, determined by the   transverse equations \cite{DM-2}, 
\beq
\dfrac {d^2\!X^{\pm}}{dU^2}\mp\frac{1}{2}\cA(U) X^{\pm} = 0\,,
\label{geoX1X2}
\eeq
where the $\pm$ label refers to the upper resp. lower component.
The sign of the second\; term comes from the
 relative minus between the coordinates in \eqref{Bmetric}.
These Sturm-Liouville-type equations are completed with the DM boundary conditions,
\beq
\frac{d\bX}{dU}\Big|(U=-\infty) = 0 = \frac{d\bX}{dU}\Big|(U=+\infty)\,.
\label{DMcond}
\eeq

Our main result then says that for the profile \eqref{Scarfpot} {analytic solutions can be found}, consistently with the
prediction of Zel'dovich and Polnarev \cite{ZelPol}, supplemented by the quantization condition mentioned above.
Our clue is that the \emph{geodesic equation \eqref{geoX1X2} can be viewed as a time-independent Schr\"odinger equation} and the \emph{DM trajectories correspond to its real} (non-normalizable) \emph{zero-energy ground states} \cite{DM-1}.

\section{Analytic solution for flyby}\label{AnalFly}

We first recall how things go for the simple \PT case \cite{DM-1,DM-2}. Putting $t=\tanh U$ into eqn. \eqref{geoX1X2} with  profile $\cA^{PT}=(g/2)\cosh^{-2}U$  yields,
\beq
(1-t^2)\frac{d^2X^{\pm}}{dt^2}-2t\frac{dX^{\pm}}{dt}
\mp\frac{g}{4}X^{\pm}=0\,. 
\label{PTeq}
\eeq
Let us consider first one of the components, say $X^{+}$. Then DM is obtained in terms of Legendre polynomials \cite{DM-1} when the coefficient in the linear-in-$X^{+}$ term is quantized, as
\beq
g=-4n(n+1) \where n>0 \quad \text{is a natural integer}\,.
\label{PTg}
\eeq

However passing to  the other component, the last, linear-in-$X^{-}$
term has  opposite sign, therefore DM would be obtained for $-g$. But this is impossible because  $g$  has already been fixed. In conclusion, one has what we called ``half DM" \cite{DM-2}: we get a DM solution in one or in the other sector, but not for both in general --- unless the ``wrong'' component is set to zero.

Solutions for the Scarf potential $\cA^{Scarf}$ (and similarly for derived \PT\!, $\cA^{dPT}$,
 \cite{DM-2}), are obtained by
putting  $t=\sinh{U}$. Then  \eqref{geoX1X2} is written 
\begin{equation}
    (1+t^2)\frac{d^2X^{\pm}}{dt^2}+t\frac{dX^{\pm}}{dt} {\mp}g\frac{t}{1+t^2}X^{\pm}=0\,. 
    \label{eq3}
\end{equation}
The two equations here are uncoupled and can therefore be solved independently.  Note however the coefficient $t$ in front of the linear-in-$X^{\pm}$ term whose sign change compensates the $\pm$ sign change between the components. Therefore the same $g$ works for both components, as it will be seen in sec. \ref{NUgeo}.

Finding the geodesics for dPT is considerably more difficult, though, as for ``underived" \PT\,; it can be obtained in terms of confluent Heun functions \cite{DM-2} with numerically calculated approximate expressions, see eqn. \# (IV.16) of \cite{DM-2}. 
\goodbreak

Turning to Scarf, we note that \eqref{eq3}
is of the generalised hypergeometric type which can be solved analytically by the Nikiforov-Uvarov (NU) method \cite{NikiUvar}. 
Since a fairly detailed description of this simple yet elegant and powerful method can be found elsewhere \cite{NikiUvar,NU1,NU2,NU3,NU4}, here we just list the main results. The algorithm, whose details are discussed in sec.\ref{NUgeo},  yields  
the desired ``magic'' coefficient $g$ in \eqref{eq3},
\begin{equation}
|g|=|g_n|=(2n+1)\sqrt{n(n+1)}\,,
\label{eq43BIS}
\end{equation} 
 where $n$ is a natural integer.
Then the trajectory is,
 \begin{equation}
X_n(t)=(1+t^2)^{-\frac{n}{2}}\,e^{-\sign(g)\,\sqrt{n(n+1)}\arctan{t}}\, y_n(t)\,,
\label{eq47BIS}
\end{equation}  
 where 
 $y_n(t)$ is the polynomial 
\begin{equation}
y_n(t)=\frac{B_n}{\rho(t)}\left [(1+t^2)^n\rho(t)\right]^{(n)}\,
\label{eq50BIS}
\end{equation}
 with weight function 
\begin{equation}
\rho(t)=(1+t^2)^{-\left(n+\frac{1}{2}\right)} e^{-2\sign(g)\,\sqrt{n(n+1)}\arctan{t}}\,.
\label{eq49BIS}
\end{equation}

The normalization constants $B_n$ is determined from the initial conditions. 

The analytic solutions \eqref{eq47BIS} reproduce almost identically those obtained numerically in FIG.s \# 8 \& 9 of \cite{DM-2} for the derived \PT case.
 For $n\geq5$, though, the numerical procedure becomes tedious as it requires to calculate a high number of decimals. For high wave numbers the result becomes inaccurate because the procedure used in \cite{DM-2} is only an approximation. Plotting instead the analytic solution \eqref{eq47BIS}-\eqref{eq49BIS}
 yields accurate figures, underlining the advantage of the analytical approach. 

\begin{figure}[h]
\includegraphics[scale=.58]{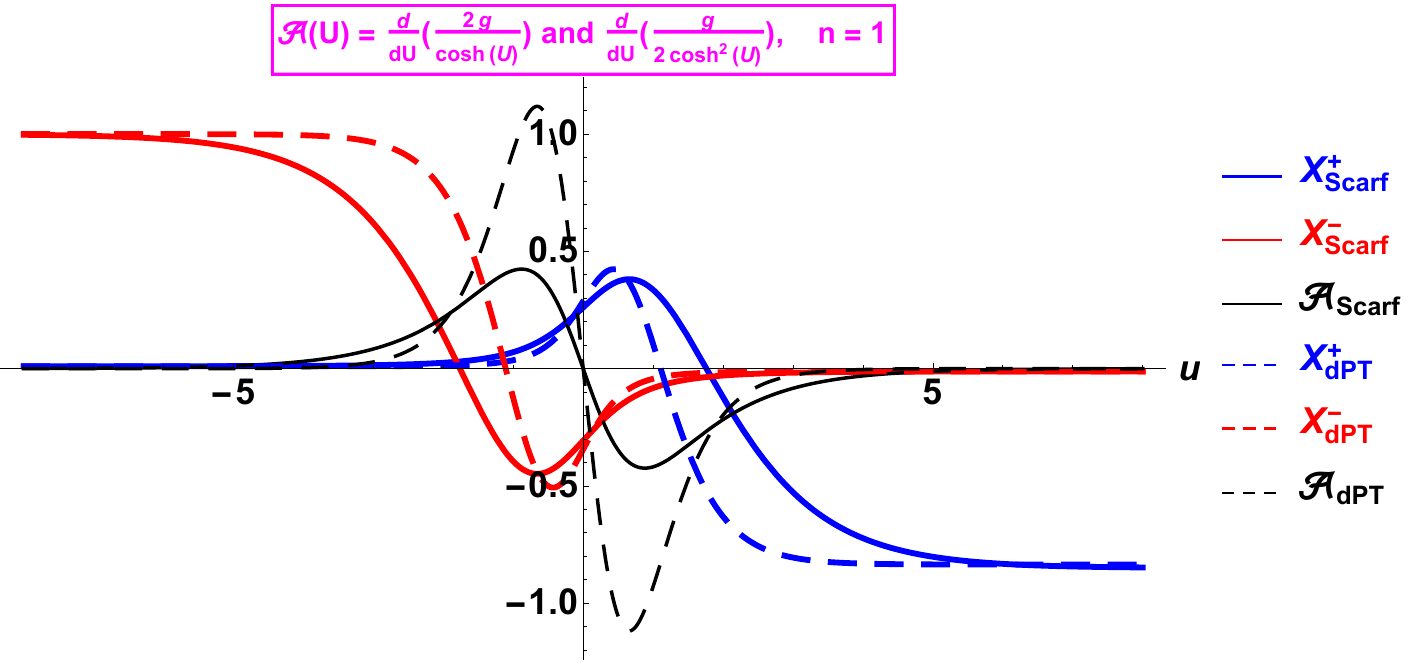}
\\[10pt]
\includegraphics[scale=.58]{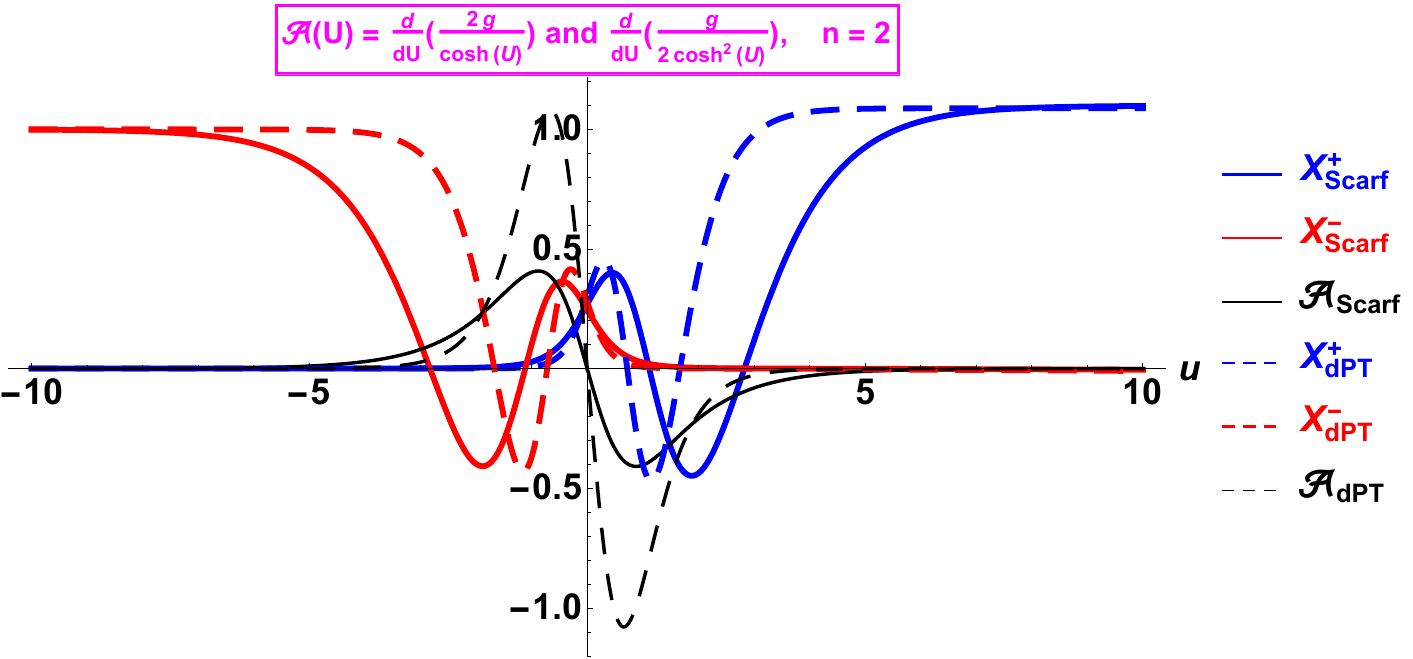}
\vskip-5mm\caption{\textit{\small The DM geodesics for the derived \PT \eqref{dPTprof} and for the Scarf profiles \eqref{Scarfpot} are similar however not fully identical, as shown for wave numbers ${\bf n=1}$ and ${\bf n=2}$. (The trajectories were rescaled for better comparison).
}
\label{dPTScarf12}
}
\end{figure}

\section{Transverse geodesics by the Nikiforov-Uvarov method}\label{NUgeo}

The Nikoforov-Uvarov method (recalled in the Appendix) applies to second-order differential equations of the generalised hypergeometric type of the form,
\begin{equation}
u^{\prime\prime}+\frac{\pi_1(z)}{\sigma(z)}u^\prime+\frac{\sigma_1(z)}{\sigma^2(z)}u=0\,.
\label{eq4} 
\end{equation}
In the Scarp case Eq. \eqref{eq3} 
$ 
\sigma=1+t^2,\;\pi_1=t,\;\sigma_1=-gt\,.
$ 
Then, 
\begin{equation}
\sigma_3=\frac{t^2}{4}+gt+k(1+t^2)=t^2\left(k+\frac{1}{4}\right)+gt+k\,.
\label{eq33}
\end{equation}
Therefore, the condition $\det(\sigma_3)
=g^2-4k\left(k+\frac{1}{4}\right)=0$ gives
\begin{equation}
k=\frac{\pm\sqrt{1+16g^2}-1}{8}
\aand
|g|=2\sqrt{k\left(k+\frac{1}{4}\right)}\,.
\label{eq35-36}
\end{equation}
Thus we have
\begin{equation}
\sigma_3=\frac{g^2}{4k}\left [t^2+\frac{4k}{g}t+\frac{4k^2}{g^2}\right]=\frac{g^2}{4k}\left(t+\frac{2k}{g}\right)^2,
\;
\pi=\frac{t}{2}\pm \sqrt{\sigma_3}= \frac{t}{2}\pm\frac{|g|}{2\sqrt{k}}\left |t+\frac{2k}{g}\right|\,.\: 
\label{eq37}
\end{equation}
Then
\begin{equation}
\tau=\pi_1+2\pi=2t\pm \frac{|g|}{\sqrt{k}}\left |t+\frac{2k}{g}\right|.
\label{eq38}
\end{equation}
To have real solutions, we assume $k>0$, that is we choose the upper sign 
$ 
k=\frac{\sqrt{1+16g^2}-1}{8}   
$ 
in Eq.\eqref{eq35-36}.
On the other side, the quantization condition Eq. \eqref{eq24} requires
\begin{equation}
\lambda= -n(n+1)\mp\epsilon\,\frac{n|g|}{\sqrt{k}}\,,
\label{eq40}
\end{equation}
since
$\tau^\prime=2\pm\epsilon\,\frac{|g|}{\sqrt{k}}$ with  
$\epsilon=\sign\left(t+\frac{2k}{g}\right)\,$
and we get, after some algebra,
$$
k=\lambda-\pi^\prime=-n^2-n-\frac{1}{2}\mp\epsilon\frac{|g|}{\sqrt{k}}\left(n+\frac{1}{2}\right ),
$$
and using Eq. \eqref{eq35-36} for $|g|/\sqrt{k}$, we get 
$ 
\sqrt{k+\frac{1}{4}}=\mp\epsilon\left(n+\frac{1}{2}\right)
$ 
and we see that we must choose the lower signs in Eq.\eqref{eq37} and Eq.\eqref{eq38} to get $\mp\epsilon=1$. Then the solution is
$ 
k_n=n(n+1)\,,
$ 
and Eq.\eqref{eq35-36} determines the ``magic" strengths labeled by the integer ``quantum'' number $n$, eqn. \eqref{eq43BIS}, for which we have polynomial solutions $y_n(t)$. 

There remains to find the gauge function $\varphi(t)$ and the corresponding hypergeometric type polynomials $y_n$.
 We have,
\begin{equation}
\pi(t)=-nt-\sign(g)\sqrt{n(n+1)}, \;\;\;\; \tau(t)=
-(2n-1)t-2\sign(g)\sqrt{n(n+1)}\,.
\label{eq44}
\end{equation}
The equation for the gauge function,
\begin{equation}
\frac{d\varphi}{dt}= -\frac{nt+\sign(g)\sqrt{n(n+1)}}{1+t^2}\,,
\label{eq45}
\end{equation}
can be easily integrated to yield (up to an irrelevant additive constant which translates to a multiplicative constant in the solution $X(t)$),
\begin{equation}
\varphi(t)=-\frac{n}{2}\ln(1+t^2)+\sign(g)\,\sqrt{n(n+1)}\arctan{t}\,.
\label{eq46}
\end{equation}
Therefore, the corresponding geodesic trajectory is given by
 eqn. \eqref{eq47BIS}.
The polynomial $y_n(t)$ is given by the Rodrigues formula Eq. \eqref{eq31}, which requires to calculate the weight function $\rho(t)$ in  \eqref{eq49BIS}. 
Then we end up with eqn. \eqref{eq50BIS}.

Eqn.  \eqref{eq49BIS} implies that $\rho(t)$ does not change when $t$ and $g$ change signs simultaneously, while due to the derivatives in eqn. \eqref{eq50BIS}, the polynomials $y_n(t)$ acquire a $(-1)^n$ factor. Then eqn. \eqref{eq47BIS} implies the parity behaviour
\beq
X^{\pm}_n(-t) = (-1)^n X^{\mp}_n(t)\,
\label{parity}
\eeq
mentioned in sec.\ref{AnalFly}.
Numerical evidence suggests that $t=0$ is not a root of $X^{\pm}_n$ neither for dPT nor for Scarf and, in fact, for a profile which is an order-odd derivative of an even function. On the contrary, it is a root when the profile is an order-even derivative.

The curious pairing of the components in \eqref{parity}
is reminiscent of and is indeed closely related to 
\emph{supersymmetry} \cite{MSUSY}. 

The $y_n(t)$ can be expressed in terms of Routh-Romanovski polynomials \cite{Scarfpot,Raposo} as $y_n(t)=R_n^{(\alpha,\beta)}(t)$ with $\alpha=-2\,\sign(g)\sqrt{n(n+1)}$ and $\beta=-n+\frac{1}{2}$. The first three of them are  
$
y_1(t)\sim 
-t-2\sqrt{2}\,\sign(g),
\; y_2(t)\sim 2t^2+8\sqrt{6}\,\sign(g) t+23, 
\;y_3(t)\sim -6t^3-72\sqrt{3}\,\sign(g)t^2-423t-172\sqrt{3}\,\sign(g)\,.
$
\section{Conclusion}\label{Concl}

The toy profile \eqref{dGprof} proposed in \cite{LongMemory}, which may or may not be similar to that of flyby encountered in nature, yields, for certain ``magic'' values of the wave parameter, a step-like, pure displacement behavior, consistently with the suggestions of Zel'dovich and Polnarev.
The profile \eqref{dGprof} can, moreover,   
 be approximated by both the derived-\PT \eqref{dPTprof} and by the Scarf potentials \eqref{Scarfpot}. The latter supports  analytic solutions,  
\eqref{eq47BIS}-\eqref{eq50BIS}, constructed by the Nikiforov-Uvarov method and  illustrated in FIG.\ref{ScarfProfFig}. The solutions are labeled by an integer $n$, cf. eqn \eqref{eq43BIS}.

The  displacement effect (DM) arises when the wave zone contains $n$ half-waves as observed in  \cite{DM-2} for the Gaussian and \PT profiles. The number of half-waves is equal to that of the zeros of the polynomial $y_n(t)$. 

Our Scarf solution is similar to but slightly different from those found semi-analytically for derived-\PT \cite{DM-2}  FIG.\ref{dPTScarf12}. It is exact, while the  one in \cite{DM-2} is only approximate, with error increasing for higher wave numbers.

The  behavior of both the Scarf \eqref{Scarfpot} and of the derived-\PT \eqref{dPTprof} is explained by their parity behaviour, \eqref{parity}, which explains in turn why do we get DM for both components. The same is true for any odd-order derivative of an even-profile function  \cite{DM-1,DM-2,Benin}. For even-order derivatives the analogous choices of the parameters  would be contradictory, implying that one can gets DM only for one, but not for both components --- unless the other is put to zero. 

\goodbreak

Physical applications of various incarnations of the Scarf potential are reviewed in ref. \cite{Raposo}. They include  supersymmetry \cite{Scarfpot,Scrf1,Compean,Gendenshtein,MSUSY} and non-central potentials \cite{Dutt86}. 
The recent paper \cite{MSUSY}, which is directly relevant for our purposes here, highlights the prominent role of SUSY QM for \GWs with Scarf profile. See also \cite{Benin,Hait,Bhattacharya23,Bhattacharya:2025ljc} for further related work. 

We mention that our main tool here, namely the Nikiforov-Uvarov method, applies also to a number of other singular and periodic GW profiles, such as
\begin{equation}
{\cal A}=2g\left(1-\frac{1}{\tanh^2{U}}\right),\;\:{\cal A}=-2g\cn^2{\left(U,\frac{1}{\sqrt{2}}\right)},\;\;{\cal A}=2g\tan{U},\;\;{\cal A}=\frac{2g}{\tan{U}}\,,
\label{OtherAs}
\end{equation}
where $\cn(z,k)$ is Jacobian elliptic cosine function. Interestingly, the last profile corresponds to the Coulomb potential in constant curvature space $\mathbb{S}^3$, and hydrogenlike atoms in constant curvature spaces are most naturally considered by the Nikiforov-Uvarov method \cite{Alizzi}.

It has long been known (see, e.g., 
\cite{Gendenshtein,Dutt86, Scarfpot,Cooper94,Dutt92})
 that the \PT and (hyperbolic) Scarf potentials are parts of the same supersymmetric partner potentials, following from the 
shape invariant superpotential
\beq
W(U)=A{\,}\tanh(U)+B{\,}{\rm sech}(U)\,.
\label{DuttPot}
\eeq
 A recent paper \cite{MSUSY}   confirms our results 
using  supersymmetric quantum mechanics.

\goodbreak


\acknowledgements{
The authors are grateful to Gary Gibbons and to M. Elbistan for discussions and to G. Natanson for correspondance. PMZ was partially supported by the National Natural Science Foundation of China (Grant No. 11975320).
}

\goodbreak



\appendix
\section{\bf Compendium on the Nikoforov-Uvarov method
}\label{Appendix}

As it was already mentioned in the text, the Nikoforov-Uvarov method applies to second-order differential equations of the generalised hypergeometric type of the form
\begin{equation}
u^{\prime\prime}+\frac{\pi_1(z)}{\sigma(z)}u^\prime+\frac{\sigma_1(z)}{\sigma^2(z)}u=0\,,
\label{eq4bis} 
\end{equation}
where the prime denotes $d/dz$ where $z$ can be complex. $\pi_1(z)$ is a polynomial at most of the first degree, and $\sigma(z)$, $\sigma_1(z)$ are polynomials at most of the second degree \cite{NikiUvar}. 

The set of solutions of Eq. \eqref{eq4bis} is invariant under  ``gauge" transformations $u(z) \to y(z)$,
$ 
u(z)=e^{\varphi(z)}y(z),    
$ 
if the gauge function $\varphi(z)$ satisfies the equation
\begin{equation}
\varphi^\prime(z) =\frac{\pi(z)}{\sigma(z)}\,,
\label{eq7}
\end{equation}
where $\pi(z)$ is a polynomial at most of the first degree. In this case for $y(z)$ 
we obtain also a generalised hypergeometric-type equation:
\begin{equation}
y^{\prime\prime}+\frac{\pi_2(z)}{\sigma(z)}y^\prime+\frac{\sigma_2(z)}{\sigma^2(z)}y=0\,,
\label{eq8}    
\end{equation}
where
$
\pi_2(z)=\pi_1(z)+2\pi(z)
$ 
is a polynomial  of at most of the first degree, and
\begin{equation}
\sigma_2(z)=\sigma_1(z)+\pi^2(z)+\pi(z)\left [\pi_1(z)-\sigma^\prime(z)\right]+\pi^\prime(z)\sigma(z)
\label{eq10}
\end{equation}
is a polynomial of at most of the second degree. 
Choosing  $\pi(z)$ so that
\begin{equation}
\sigma_2(z)=\lambda\sigma(z)\,   
\label{eq11}
\end{equation}
where $\lambda$ is a constant, Eq. \eqref{eq4bis} is simplified to an hypergeometric type equation,
\begin{equation}
\sigma(z)y^{\prime\prime}+\pi_2(z)y^\prime+\lambda y=0\,.
\label{eq12}
\end{equation}

In the light of Eq. \eqref{eq10}, Eq. \eqref{eq11} means
\begin{equation}
 \pi^2+\pi[\pi_1-\sigma^\prime]+\sigma_1-k\sigma=0,
 \label{eq13}
\end{equation}
where
$
k=\lambda-\pi^\prime
$ 
is another constant.
Eq. \eqref{eq13} is  solved by,
\begin{equation}
\pi=\frac{\sigma^\prime-\pi_1}{2}\pm\sqrt{\sigma_3(z)}
\where
\sigma_3(z)=\left(\frac{\sigma^\prime-\pi_1}{2}\right)^2-\sigma_1+k\sigma\,.
\label{eq15}
\end{equation}
$\pi$ is a polynomial only when
$\sigma_3(z)$ is the square of a first order polynomial. Hence it has a double root and its discriminant is zero,
$
\Delta(\sigma_3)=0\,,
$
which determines the constant $k$, and hence 
the constant $\lambda$.

We want to find polynomial solutions of Eq. \eqref{eq12}. It can be shown that derivatives of $y(z)$, $v_n(z)=y^{(n)}(z)$, are also generalised hypergeometric-type functions,
\begin{equation}
\sigma v_n^{\prime\prime}+\tau_n(z)v_n^\prime +\mu_nv_n=0\,,
\label{eq20}
\end{equation}
and we get the recurrence relations 
\begin{equation}
\tau_n(z)=\sigma^\prime(z)+\tau_{n-1}(z),\;\;\;\;\mu_n=\mu_{n-1}+\tau_{n-1}^\prime\,,
\label{eq21}
\end{equation}
with initial values
$\tau(z)=\tau_0(z)=\pi_2(z),\;\mu_0=\lambda\,.
$ 
If $y(z)=y_n(z)$ is a polynomial of order $n$, then $v_n=\mathrm{const}$ and  Eq. \eqref{eq20} is satisfied only if $\mu_n=0$. But the repeated application of the recurrence relations Eq. \eqref{eq21}  produces
\begin{equation}
\tau_n(z)=n\sigma^\prime(z)+\tau(z),\;\;\;\mu_n=\lambda+n\tau^\prime+\frac{1}{2}n(n-1)\sigma^{\prime\prime}\,.
\label{eq23}
\end{equation}
Getting a polynomial solution $y(z)=y_n(z)$ of Eq. \eqref{eq12} 
thus requires the ``quantization condition'',
\begin{equation}
\lambda=\lambda_n=- n\tau^\prime-\frac{1}{2}n(n-1)\sigma^{\prime\prime}.
\label{eq24}
\end{equation}
It can be shown \cite{NikiUvar} that, when the quantization condition Eq. \eqref{eq24} is satisfied, the corresponding polynomial solution of Eq. \eqref{eq12} is given by the Rodrigues formula
\begin{equation}
y_n(z)=\frac{B_n}{\rho(z)}\left[\sigma^n(z)\rho(z)\right]^{(n)},
\label{eq31}
\end{equation}
where $B_n$ is some (normalization) constant, and the weight function $\rho(z)$ satisfies Pearson's equation \cite{Ismail}  
\begin{equation}
(\sigma\rho)^\prime=\rho\tau\,.
\label{eq25}
\end{equation}

%

\end{document}